\documentclass[preprint2]{aastex}

\newcommand{\myemail}{john@physics.uwa.edu.au}

\shorttitle{Unknown selection effect in SDSS quasar number counts}
\shortauthors{Hartnett}

\begin{document}

\title{Unknown selection effect simulates redshift periodicity in quasar number counts from Sloan Digital Sky Survey}

\author{J.G. Hartnett}
\affil{School of Physics, the University of Western Australia\\35 Stirling Hwy, Crawley 6009 WA Australia; \myemail}

\begin{abstract}
Discrete Fourier analysis on the quasar number count, as a function of redshift, $z$, calculated from the Sloan Digital Sky Survey DR6 release  appears to indicate that quasars have preferred periodic redshifts with redshift intervals of 0.258, 0.312, 0.44, 0.63, and 1.1. However the same periods are found in the mean of the $zConf$ parameter used to flag the reliability of the spectroscopic measurements. It follows that these redshift periods must result from some selection effect, as yet undetermined. It does not signal any intrinsic (quantized) redshifts in the quasars.
\end{abstract}


\maketitle  

\section{\label{sec:Intro}Introduction}
Over the past three decades or more astronomical evidence has been interpreted by some as suggesting that a major component of the redshift of quasars arises from a non-cosmological origin \citep{Burbidge1968, Karlsson1971, Karlsson1977, Burbidge1990, Burbidge2001, Bell2002a, Bell2002b, Bell2002c}, which is quantized in some fashion. They offered schemes of quantization of the redshift, $z$, either linear in $z$ or as a function of $log(1+z)$. It has also been suggested that the intrinsic component of redshift is some harmonic of a preferred value. Amongst the earliest reports \citep{Burbidge1968} found that $z = 0.061$ occurred frequently suggesting something intrinsic. \citep{Burbidge1968} also suggested a quantization scheme of the form $0.061 n$ where $n$ is an integer. Then more recently \citep{Bell2002b} suggested that all quasar redshifts contain some harmonic of $0.062$. 

Recently \citep{Bell2006} analyzed the data from the third data release of the the Sloan Digital Sky Survey \citep{SDSS} and found a significant peak in the power spectrum near $\Delta z = 0.62$. Here $\Delta z$ represents the periodic interval seen in quasar redshift abundances. In this paper I analyze the SDSS sixth quasar data release using a Fourier transform of their redshift abundances as a function of redshift. I show, regardless of any interpretation of the meaning of the redshifts, and aside from any cosmological assumptions, that there is a significant periodicity in the SDSS quasar redshift abundance data. But also the same periods are seen in the mean of the parameter $zConf$ supplied by the Sloan survey people to flag reliability of the redshift. Moreover if one filters the data by the $zConf$ parameter the peaks in the Fourier spectrum are enhanced. This proves that, whatever their cause, they are fundamentally connected to the algorithm that assigned the $zConf$ parameter and not to some intrinsic effect from the quasars.

\section{$N(z)$ relation}
I obtained  80,398 quasar data from the SDSS BestDR6 database \citep{Adelman-McCarthy} available at $\verb"cas.sdss.org/astro/en/"$.  The data were selected by $\verb"specClass" = 3$ for $\verb"QSO"$ and $4$ for $\verb"HIZ_QSO"$, low redshift and high redshift quasars, respectively. These include any objects with spectra that have been classified by the spectroscopic pipeline as quasars ($\verb"specClass" = \verb"QSO"$ or $\verb"HIZ_QSO"$).

\begin{figure}
\includegraphics[width = 3 in]{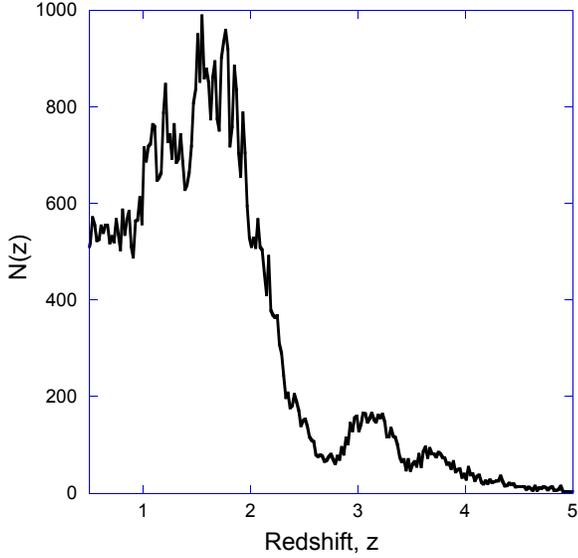}
\caption{\label{fig:fig1} The $N(z)$ relation with redshift bins $\delta z = 2\times 10^{-2}$, for the SDSS DR6 quasar data used here $0.5 < z < 5$.}
\end{figure}

The DR6 data used here were not filtered as was the DR5 quasar catalog, described in Schneider et al. \citep{Schneider}. In that case the DR5 catalog quasars were chosen from those that have apparent i-band PSF magnitudes fainter than 15, absolute i-band magnitudes brighter than -22, contain at least one emission line or are unambiguously broad absorption line quasars, and have highly reliable redshifts. In the latter cosmological assumptions were required to obtain absolute magnitudes. For this analysis such assumptions were avoided. However to exclude low redshift misidentification, and noise from insufficient high redshift data, I used only data where $0.5 < z < 5$.  The $N(z)$ relation was calculated by binning the number of quasars with redshifts that fall between $z$ and $z + \delta z$. The resulting $N(z)$ plot is shown in fig. \ref{fig:fig1}.   

\begin{figure}
\includegraphics[width = 3 in]{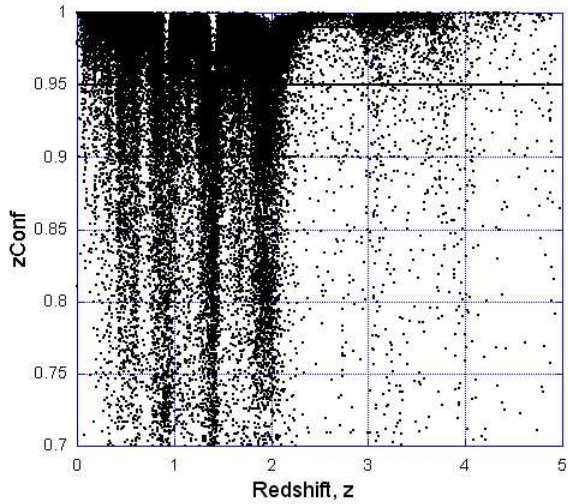}
\caption{\label{fig:fig2} The parameter $zConf$ as a function of redshift, $z$.}
\end{figure}

Quasars are identified optically and their redshifts determined from their bright emission lines. The parameter $zConf$ was assigned as figure of merit of the confidence on the redshift. It is plotted here as a function of redshift, $z$, in fig. \ref{fig:fig2}. 

A possible correlation between the under abundances, or valleys in the $N(z)$ function, at certain redshifts has been studied by \citep{Bell2006}.  They found little likelihood of any selection effect on the data they used (DR3 in their case). See their fig. 11 and the discussion thereabout. The valleys in  fig. \ref{fig:fig1} are found in the same redshift regions and the abundances shown  generally agree with the 6 peaks observed by \citep{Bell2006}.

\subsection{Fourier Spectra}
\begin{figure}
\includegraphics[width = 3 in]{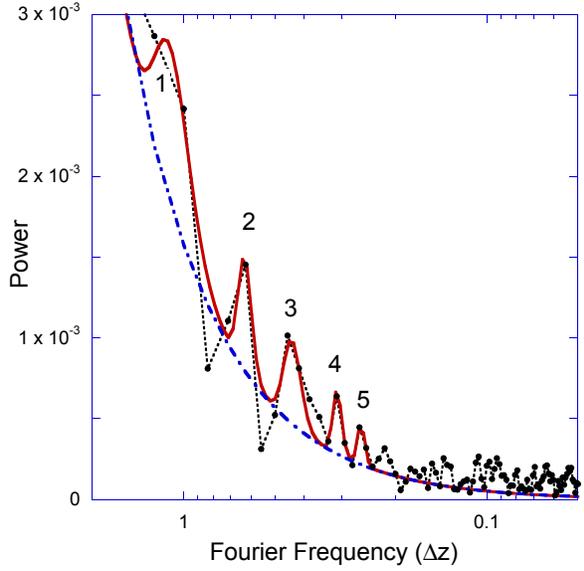}
\caption{\label{fig:fig3} The power spectrum  calculated from $N(z)$ resulting from the SDSS DR6 quasar  data binned with $\delta z = 2\times 10^{-3}$ (filled circles, dotted line). The solid (red) curve here is the sum of 5 Gaussian curves  fitted to the spectrum. The dot-dashed (blue) line is a power law fit neglecting any peaks.}
\end{figure}

\begin{figure}
\includegraphics[width = 3 in]{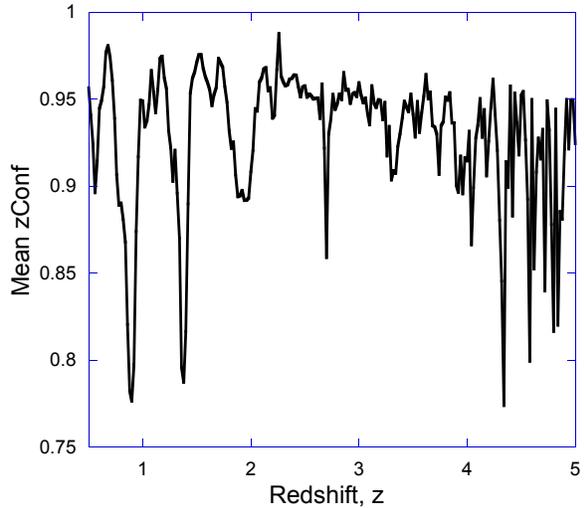}
\caption{\label{fig:fig4} The mean value of the quasar $zConf$ parameter binned with $\delta z = 2\times 10^{-2}$ and the  plotted as a function of redshift $0.5<z<5$.}
\end{figure}

The discrete Fourier Transforms were calculated with Mathematica software from the $N(z)$ relation determined by binning $z$ (fig. \ref{fig:fig3}) and from the mean value of $zConf$ binned with $\delta z = 2\times 10^{-2}$ (fig. \ref{fig:fig4}). For the Fourier spectrum shown in fig. \ref{fig:fig3} the quasar redshift, $z$, data were binned with $\delta z = 2\times 10^{-3}$. The peaks are labeled 1 through 5.    

Within the $i$th bin there are $\Delta N_i$ samples. Discrete Fourier amplitudes $\nu_s$ are generated for each integer $s>1$ from
\begin{equation} \label{FFT}
\nu_s =\frac{1}{\sqrt{n}}\sum_{i=1}^n u_s e^{2\pi j(i-1)(s-1)/n},
\end{equation}
where $u_s = \Delta N_i$, $j=\sqrt{-1}$, $n$ is the total number of $\Delta N_i$ bins.  Because $\nu_s$ are complex their absolute value is taken in the analysis. The Fourier frequencies are calculated from
\begin{equation} \label{Freq}
\frac{s}{n \delta z} =\frac{1}{\Delta z}.
\end{equation}
where $\Delta z$ is the redshift interval of any periodic structure in redshift space. In order to more easily identify the redshift interval in the power spectra used here I have converted the Fourier frequency axis to units of $\Delta z$.


The power spectrum from $N(z)$ was then fitted with the sum of 5 Gaussian curves, shown in fig. \ref{fig:fig3} as the solid (red) curve. The resulting best fit peak $\Delta z$ values and their standard deviations are listed in Table I. From this figure it is apparent that the peaks are significant to about 3$\sigma$ when compared to the higher Fourier frequency noise. In order to obtain the best fits I first fitted a power law to the data of fig. \ref{fig:fig3} while ignoring the peaks. This is shown by the dot-dashed (blue) line, which was subsequently used with the Gaussian fit. The fit to peak number 1 is very dependent on the power law fit, so its peak value has the largest uncertainty.  Peak number 2 is the same as that found by \citep{Bell2006}. 

\begin{table}[ph]
\begin{center}
Table I: Redshift $\Delta z$ intervals\\
\vspace{6pt}
\begin{tabular}{c|cc} \hline
\hline
Peak No 		&$\Delta z$ 	&error 	\\
\hline
\hline
1			&1.1					&0.2  \\
2			&0.63					&0.05 	\\
3			&0.44					&0.04 \\
4 		&0.312				&0.014 \\
5			&0.258				&0.010 \\
\hline
\hline
\end{tabular}
\end{center}
\end{table}

\begin{figure}
\includegraphics[width = 3 in]{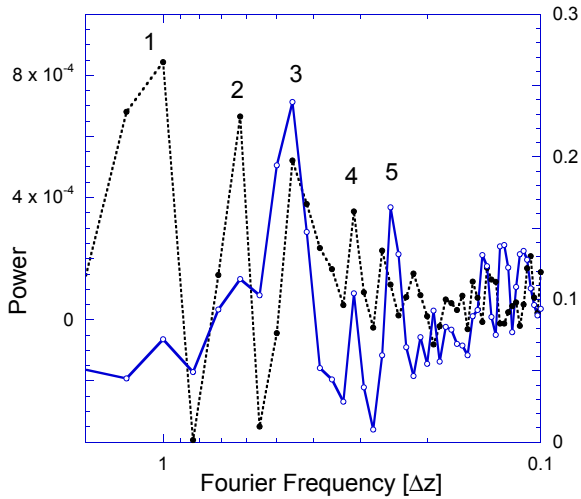}
\caption{\label{fig:fig5} The power spectrum  of the SDSS DR6 quasar $N(z)$ data with power law dependence removed (filled circles, black dotted line) with the power spectrum  of the mean of the $zConf$ parameter (open circles, solid (blue) line) on a double Y-plot. }
\end{figure}

The mean of the $zConf$ parameter binned with $\delta z = 2\times 10^{-2}$ is shown in fig. \ref{fig:fig4}. The Fourier Transform of this data was then taken and the result is shown in fig. \ref{fig:fig5}, compared with the Fourier spectrum of the $N(z)$ data of the quasar redshifts drawn from fig. \ref{fig:fig3} but with the power law dependence (dot-dashed (blue) curve) removed. The numbered peaks correspond to the same peaks of fig. \ref{fig:fig3}.

\begin{figure}
\includegraphics[width = 3 in]{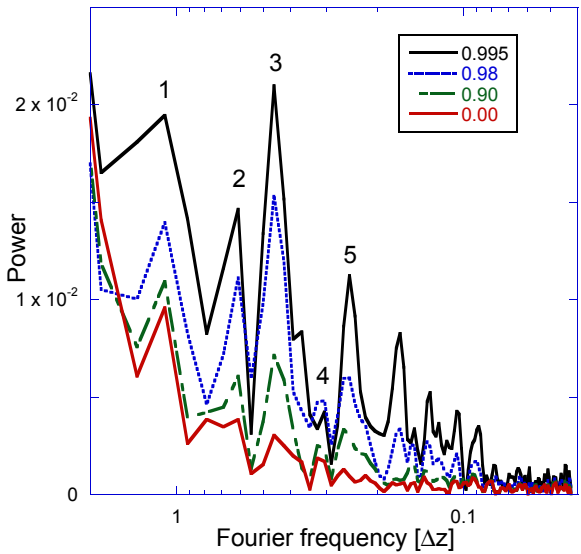}
\caption{\label{fig:fig6} The power spectrum  of the SDSS DR6 quasar $N(z)$ data where the redshift data are filtered by $zConf$ parameter (indicated in the legend) before $N(z)$ were calculated and the Fourier spectrum taken. $zConf = 0.00$ indicates no cut was taken, all the data were used.}
\end{figure}

\section{Discussion}
It is quite obvious from fig. \ref{fig:fig5} that the periodic structure  of the mean of the $zConf$ parameter in redshift space corresponds exactly with the periodic structure of the redshift abundances themselves, indicated by the coincidence of the peaks in their Fourier power spectra.  Therefore this suggests that the origin of any periodic redshift dependence in the quasars is an artifact of the sampling algorithm. 

This argument is further strengthened when we look at the Fourier spectra of the $N(z)$ function  where the redshift data have been filtered by the $zConf$ parameter. See fig. \ref{fig:fig6}. As cuts in the data of higher and higher values of $zConf$ are taken stronger Fourier peaks emerge. This indicates also a strong correlation between the two. It is worth noting that the indicated peaks, corresponding to the redshift periods $\Delta z$ of Table I, can be approximated by $0.062n$ where the integer $n = 4, 5, 7, 10$ and $20$, within the standard errors of their Gaussian fits. One could speculate that this implies an intrinsic redshift \citep{Bell2002c}, quantized in some fashion, but this analysis seems to indicate otherwise.

\section{Conclusion}
Fourier spectral analysis has been carried out on the abundance of quasar redshifts as a function of redshift from the SDSS DR6 data release. The analysis finds that there are preferred redshifts separated by intervals of $\Delta z = $0.258, 0.312, 0.44, 0.63, and 1.1, which could all be harmonics of some more fundamental value $\Delta z = 0.062$. However there is also a strong correlation with the $zConf$ parameter used to flag the quality of the $z$ parameter itself. Regardless of the algorithm used to determine $zConf$, this fact then strongly brings into doubt any validity to the argument that these periodic redshifts indicate preferred intrinsic redshifts for the quasars in the Sloan survey. Most probably an as-yet-undetermined selection effect simulates these periodic preferred redshifts. 

\section{Acknowledgments}
I would like to thank F. Oliveira for obtaining the SDSS data for me. This work has been supported by the Australian Research Council.

Funding for the SDSS and SDSS-II has been provided by the Alfred P. Sloan Foundation, the Participating Institutions, the National Science Foundation, the U.S. Department of Energy, the National Aeronautics and Space Administration, the Japanese Monbukagakusho, the Max Planck Society, and the Higher Education Funding Council for England. 

The SDSS is managed by the Astrophysical Research Consortium for the Participating Institutions. The Participating Institutions are the American Museum of Natural History, Astrophysical Institute Potsdam, University of Basel, University of Cambridge, Case Western Reserve University, University of Chicago, Drexel University, Fermilab, the Institute for Advanced Study, the Japan Participation Group, Johns Hopkins University, the Joint Institute for Nuclear Astrophysics, the Kavli Institute for Particle Astrophysics and Cosmology, the Korean Scientist Group, the Chinese Academy of Sciences (LAMOST), Los Alamos National Laboratory, the Max-Planck-Institute for Astronomy (MPIA), the Max-Planck-Institute for Astrophysics (MPA), New Mexico State University, Ohio State University, University of Pittsburgh, University of Portsmouth, Princeton University, the United States Naval Observatory, and the University of Washington.

\end{document}